\documentclass[
 reprint,
superscriptaddress,
 amsmath,amssymb,
 aps,
 prl,
]{revtex4-2}

\usepackage{amsmath}

\usepackage{graphicx}
\usepackage{dcolumn}
\usepackage{bm}
\usepackage{siunitx}

\begin{document}

\preprint{}

\title{Topology of Nonequilibrium Currents Controls Active Transport}

\author{Arin Escobar}
\affiliation{Department of Theoretical Condensed Matter Physics, Universidad Autonoma de Madrid, 28049, Madrid, Spain}
\affiliation{Condensed Matter Physics Center (IFIMAC), Universidad Autonoma de Madrid, 28049, Madrid, Spain}
\affiliation{Instituto Nicolas Cabrera, Universidad Autonoma de Madrid, 28049, Madrid, Spain}

\author{Galor Geva}
\affiliation{Department of Theoretical Condensed Matter Physics, Universidad Autonoma de Madrid, 28049, Madrid, Spain}
\affiliation{Condensed Matter Physics Center (IFIMAC), Universidad Autonoma de Madrid, 28049, Madrid, Spain}
\affiliation{Instituto Nicolas Cabrera, Universidad Autonoma de Madrid, 28049, Madrid, Spain}

\author{Alfredo Alexander-Katz}
\affiliation{Department of Materials Science and Engineering, Massachusetts Institute of Technology, Cambridge, MA, 02139, USA}

\author{Laura R. Arriaga}%
\affiliation{Department of Theoretical Condensed Matter Physics, Universidad Autonoma de Madrid, 28049, Madrid, Spain}
\affiliation{Condensed Matter Physics Center (IFIMAC), Universidad Autonoma de Madrid, 28049, Madrid, Spain}
\affiliation{Instituto Nicolas Cabrera, Universidad Autonoma de Madrid, 28049, Madrid, Spain}

\author{J.V. Alvarez}
\email{jv.alvarez@uam.es}
\affiliation{Department of Condensed Matter Physics, Universidad Autonoma de Madrid, 28049, Madrid, Spain}
\affiliation{Condensed Matter Physics Center (IFIMAC), Universidad Autonoma de Madrid, 28049, Madrid, Spain}
\affiliation{Instituto Nicolas Cabrera, Universidad Autonoma de Madrid, 28049, Madrid, Spain}

\author{Juan L. Aragones}
\email{juan.aragones@uam.es}
\affiliation{Department of Theoretical Condensed Matter Physics, Universidad Autonoma de Madrid, 28049, Madrid, Spain}
\affiliation{Condensed Matter Physics Center (IFIMAC), Universidad Autonoma de Madrid, 28049, Madrid, Spain}
\affiliation{Instituto Nicolas Cabrera, Universidad Autonoma de Madrid, 28049, Madrid, Spain}

\date{\today}

\begin{abstract}
Structured environments repeatedly redirect active particles, producing transport pathways that cannot be readily inferred from individual trajectories. Here, we show that the large-scale organization of these transport pathways is governed by topological constraints. Hydrodynamic scattering generates nonequilibrium current fields whose defect structure, characterized by integer indices, constrain transport pathways and renders them robust to smooth perturbations. This principle is demonstrated with rotating colloids in obstacle arrays and extended to stokeslet and force-dipole flows, thereby linking microscale transport to the topology of hydrodynamically generated nonequilibrium currents.
\end{abstract}


\maketitle

Harnessing the directed transport of microscopic particles is a central challenge in active matter and soft materials, with implications for microfluidic design~\cite{C8LC01323C}, bacterial flow management, and microrobots for targeted delivery~\cite{gao2012cargo,palagi_bioinspired_2018,bozuyuk2024roadmap}. At low Reynolds number, transport is shaped by viscous dissipation, thermal fluctuations, and boundary-mediated hydrodynamic interactions~\cite{lauga2009,spagnolie2012}. In structured environments, such as porous media and microfabricated lattices, boundaries scatter the flows generated by active particles, redirecting their motion and reshaping the accessible transport pathways~\cite{Gomez-Solano2016,dehkharghani2023,alonso-matilla2019,makarchuk2019}. Although often viewed as a source of disorder or confinement~\cite{reichhardt2014,moore2023}, this scattering may instead support robust transport through organizing principles that survive fluctuations and geometric perturbations~\cite{Gomez-Solano2016,bowick2022}. Because the resulting effective dynamics are non-conservative and break detailed balance, the problem also connects to the broader physics of nonreciprocal systems far from equilibrium~\cite{fruchart2026nonreciprocal}.

Topology provides a natural language for such robustness. In condensed matter and metamaterials, topological invariants protect edge modes and transport channels against smooth perturbations~\cite{hasan2010colloquium,kane2013,susstrunk015,souslov2017}, an idea also used to guide colloids on patterned magnetic substrates~\cite{delasheras}. In active matter, topology organizes phenomena ranging from defects in active nematics to topological modes in active metamaterials~\cite{vanzuiden2016,shankar2022topological,bowick2022}. Beyond band topology, vector-field topology classifies singular points where a direction field becomes undefined, together with their winding numbers, separatrices, and global index constraints~\cite{delpino2025,steuernagel2013,suntopology}. Fluid mechanics similarly uses streamline topology in which vortices, saddles, stagnation points, and separatrices form a dynamical skeleton governing mixing, trapping, and transport~\cite{brons2007}. In structured active fluids, however, transport is organized not simply by the solvent streamline pattern, but by the effective probability currents and orientation fields generated when the environment scatters active hydrodynamic flows. Whether these effective transport fields obey analogous topological constraints remains largely unexplored.

In this Letter, we show that structured environments organize active transport through the topology of nonequilibrium current fields. Hydrodynamic scattering introduces non-conservative components into the effective dynamics, breaking detailed balance and generating steady probability currents in real position--orientation space. Defects in the corresponding direction fields carry integer winding numbers that constrain transport pathways: noise, weak disorder, and smooth geometric perturbations can displace these defects, but cannot eliminate them except through creation, annihilation, or boundary crossing. We demonstrate this principle first in a minimal rotlet system, a rotating magnetic colloid moving within  obstacle array, and then extend it to translating hydrodynamic singularities, including stokeslets and force dipoles. Together, these results identify the topology of hydrodynamically generated currents as an organizing principle for robust active transport in structured media.

\begin{figure}[!ht]
\includegraphics[width=1\linewidth]{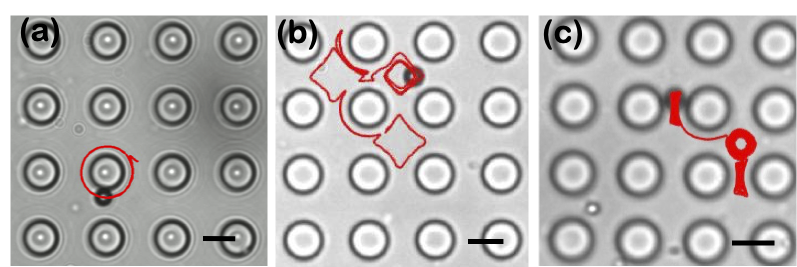}
\caption{ Optical microscope images of individual spinners of radius $a = ~$\SI{5}{\micro\meter} driven in obstacle lattices at 20~Hz in (a) and 30~Hz in (b, c) with center-to-center spacing $l_p =$ (a) \SI{21}{\micro\meter},  (b) \SI{19}{\micro\meter} and (c) \SI{17}{\micro\meter}. The circular features forming the regular array are the obstacles, while the dark particle is the spinner; solid lines show it trajectory. 
Scale bars are \SI{10}{\micro\meter}.}
\label{fig1}
\end{figure}

Our minimal experimental realization of a rotlet system consists of a dilute suspension of magnetic spheres of radius $a$ driven synchronously by a rotating magnetic field. Each particle, hereafter termed a spinner, rotates about $\hat{\mathbf z}$ with angular velocity $\Omega$, exerting a torque on the fluid and generating a rotlet flow. The structured medium is a square lattice of $N_{obs}$ cylindrical obstacles of radius $r_o$ and center-to-center spacing $l_p$, whose axes are aligned with the rotation axis. Details on the experiments can be found in the Supplementary Material (SM). Because an isolated spinner does not translate in an unbounded fluid, transport arises from obstacle-induced scattering of the rotlet flow and the resulting hydrodynamic back-action. Previous lattice--Boltzmann simulations and a coarse-grained Langevin theory showed that this mechanism converts rotation into translation, while radial lift and attraction determine spinner localization~\cite{Puerto2026}. Here, the obstacles are deliberately left unfunctionalized to avoid spinner--obstacle attraction, weakening radial localization and isolating the scattered-rotlet drift from strong trapping. Consequently, the rotation frequency controls the circulation rate rather than selecting a unique orbit. Spinners exhibit long-lived trajectories around individual obstacles, the central voids of unit cells and inter-obstacle gaps [Fig.~\ref{fig1}a--c and Supporting Movies SM1--4]. Because these circulation centers do not necessarily coincide with material obstacles, transport cannot be described as a collection of individual spinner--obstacle orbits, but is instead organized by a lattice-generated current field.

To define this organizing field, we adopt an ensemble description in terms of a probability current. The probability density $\rho(\mathbf r,t)$ obeys the overdamped Fokker--Planck equation,
\begin{equation}
    \frac{\partial \rho}{\partial t}
    =
    -\nabla\cdot\mathbf j,
    \qquad
    \mathbf j
    =
    \mathbf V(\mathbf r)\rho
    -
    D_t\nabla\rho ,
    \label{eq:fp_current_general}
\end{equation}
where $\mathbf V(\mathbf r)$ is the deterministic drift and $D_t$ is the effective translational diffusion coefficient associated with Gaussian white noise. In the long-time limit, the system reaches a stationary state $\rho_0$ with $\nabla\cdot\mathbf j_0=0$. At equilibrium, detailed balance imposes the stronger condition $\mathbf j_0=0$. By contrast, a non-conservative drift can break detailed balance and sustain a nonzero, divergence-free current in a nonequilibrium stationary state. This formulation separates localization, encoded in $\rho_0$, from circulation, encoded in $\mathbf j_0$.

For a rotlet, the obstacle-scattered flow acts directly as a translational drift. We denote the separation of the spinner from obstacle $i$ by $\mathbf R_i$, with $R_i=|\mathbf R_i|$, $\hat{\mathbf R}_i=\mathbf R_i/R_i$ and azimuthal direction $\hat{\boldsymbol{\varphi}}_i=\hat{\mathbf z}\times\hat{\mathbf R}_i$, whose sign reverses with the rotation chirality. To leading order, obstacle $i$ induces the drift $\mathbf W_i^{\varphi}=\beta a^4\Omega R_i^{-3}\hat{\boldsymbol{\varphi}}_i$,
where $\beta$ characterizes the strength and chirality of the rotation-induced hydrodynamic coupling. The $1/R_i^3$ decay was measured directly in lattice--Boltzmann simulations~\cite{Puerto2026}; here, only the smooth, sign-preserving azimuthal structure of the coupling is essential. For an arbitrary obstacle arrangement, the total drift follows by superposition, $\mathbf W^{\varphi}=\sum_i\mathbf W_i^{\varphi}$, or equivalently
\begin{equation}
    \mathbf W^{\varphi}(\mathbf r)
    =
    -\sum_{i=1}^{N_{\rm obs}} \nabla\times
    \mathbf a_i(\mathbf r),
    \qquad
    \mathbf a_i(\mathbf r)
    =
    -\frac{\beta a^4}{2}
    \frac{\Omega}{R_i^2}\hat{\mathbf z}.
    \label{eq:total_rotlet_drift}
\end{equation}
This identifies $\mathbf W^{\varphi}$ as the solenoidal, vector-potential component of the effective drift field in the Helmholtz decomposition~\cite{Puerto2026}. Thus, in the rotlet case $\mathbf V(\mathbf r)=\mathbf W^\varphi(\mathbf r)$ and the stationary probability current is
\begin{equation}
    \mathbf j_0
    =
    \mathbf W^{\varphi}\rho_0 - D_t\nabla\rho_0.
    \label{eq:steady_current}
\end{equation}
The first term carries the non-conservative circulation induced by the obstacle-scattered rotlet flow, while diffusion redistributes probability according to $\rho_0$. The stationary density $\rho_0$ therefore determines where probability accumulates, whereas the singularities, winding numbers, and separatrices of the direction field associated with $\mathbf j_0$ organize the allowed transport pathways.

We therefore define the current-direction field
\begin{equation}
    \hat{\mathbf c}(\mathbf r)
    =
    \frac{\mathbf j_0(\mathbf r)}
    {|\mathbf j_0(\mathbf r)|},
    \label{eq:current_direction}
\end{equation}
wherever $\mathbf j_0\neq 0$ within the accessible domain. Singularities occur at points where the current vanishes, $\mathbf j_0=0$, or regions excluded by a solid obstacle. Writing $\hat{\mathbf c}=c_x\hat{\mathbf x}+c_y\hat{\mathbf y}$ in the plane of motion, the winding number around a closed contour $\partial S$ lying in the accessible domain and not crossing a singularity is
\begin{equation}
   w(\partial  S)=\frac{1}{2\pi} \oint_{\partial  S} d \arg(\hat{\mathbf c})=\frac{1}{2\pi}\oint_{\partial  S} c_x d c_y-c_y dc_x .
\label{eq:winding}
\end{equation}
This integer measures net rotation of the current direction along the contour, in analogy with the winding-number classification of defects in ordered media~\cite{liquidcrystals1,liquidcrystals2,topoldefects}. A nonzero value gives the total topological index enclosed by the contour. This index is invariant under smooth deformations of the current field and can change only if a defect crosses the contour, defects are created or annhilated within it, or the boundary conditions of the accessible domain change. For numerical evaluation, we discretize the oriented contour $\partial S$ into segments joining neighboring points $\mathbf r_j$ and $\mathbf r_{j+1}$, with $\hat{\mathbf c}_j=\hat{\mathbf c}(\mathbf r_j)$. The winding number is then obtained by summing the signed angular increments of the direction field,
\begin{equation}
    w(\partial  S)
    =
    \frac{1}{2\pi}
    \sum_{j \in \partial  S}
    \operatorname{sign}\left([\hat{\mathbf c}_j \times \hat{\mathbf c}_{j+1}]_z \right) \operatorname{arccos}
    \left(\hat{\mathbf c}_j\cdot \hat{\mathbf c}_{j+1}
    \right).
    \label{eq:winding_num}
\end{equation}

\begin{figure}[h!]
\includegraphics[width=1.01\linewidth]{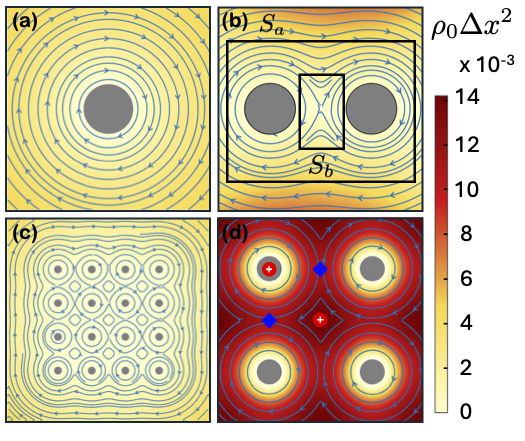}
\caption{Topological organization of steady rotlet currents in obstacle arrays.
Streamlines represent direction of the stationary probability current, and colors denote the scaled stationary density $\rho_0 \Delta x^2$, where $\Delta x$ is the grid spacing.
(a,b) One- and two-obstacle geometries, with $l_p = 2.8 r_o$ in (b). The outer contour in (b) has $w(S_a)=+1$, whereas $w(S_b)=-1$ identifies the inter-obstacle cross-defect.
(c) Finite obstacle island with $l_p=8 r_o$, where bulk cross-defects compensate obstacle-centered winding and the uncompensated circulation is carried by the edge current.
(d) Unit cell of a periodic square-lattice unit cell with $l_p=8 r_o$, showing positive circulating defects (red circles) and negative cross-defects (blue diamonds).}
\label{rotlet}
\end{figure}

For a single obstacle, the stationary current forms closed streamlines around it [Fig.~\ref{rotlet}a]. Any contour enclosing the obstacle gives $w=+1$, whereas contours that do not give $w=0$; the obstacle therefore acts as a positive-index defect with topological charge $q=1$. This topological constrain directly reflects the observed transport, as the spinner follows a nonequilibrium current that winds once around the post. The first nontrivial constraint arises with two obstacles. Although each obstacle individually contributes $w=+1$, a contour $S_a$ enclosing both gives $w=+1$, rather than $w=+2$ [Fig.~\ref{rotlet}b], requiring a compensating negative-index singularity between them. Indeed, Eq.~\eqref{eq:winding} around $S_b$ gives $w=-1$, identifying a saddle-like defect at the midpoint. We refer to such $q=-1$ singularities as \emph{cross-defects}. More generally, $w(\partial  S)=\sum_{\alpha \in S} q_\alpha$ and thus, adding obstacles does not simply superpose independent obstacle-centered vortices, but forces the current field to reorganize through compensating defects. 

Vortex-like ($q=+1$) and saddle-like ($q=-1$) singularities are the only generic isolated defects of a two-dimensional divergence-free field. Their global index balance is constrained by the Poincaré Index Theorem: on a closed surface $M$, the total index of a tangent field satisfies $\sum_{\alpha} q_\alpha=\chi(M)=2-2g$ where $g$ is the genus of $M$ and $\chi$ is its Euler characteristic. For domains with excluded obstacles or outer boundaries, the same constraint applies only after including the winding imposed by those boundaries, or equivalently by treating them as singular contributions to the compactified field (see SM). In finite obstacle islands, positive and negative defects largely compensate throughout the bulk, while the residual winding is carried by the edge current [Fig.~\ref{rotlet}c]. 
This residual index follows from the behavior of the current at large distances. The unbounded plane can be compactified by treating infinity as a single additional point; since the far-field current vanishes there while its direction winds once, infinity contributes a $q=+1$ singularity. The compactified field is topologically equivalent to the tangent field on a sphere, with $\chi=2$, and the finite defects must also carry a total index of $+1$. Accordingly, any contour $\partial S$ enclosing the obstacle island but excluding infinity has $w(\partial S)=1$. The edge current is therefore the boundary manifestation of this residual index. This relation between a bulk defect network and robust boundary transport is reminiscent of bulk--edge correspondence in topological band theory~\cite{hasan2010colloquium}, including non-Hermitian systems~\cite{nonhermitiansato}; here, however, the invariant is the real-space winding number of a nonequilibrium current field rather than a band index. 

Away from the outer edge, a sufficiently large periodic obstacle array can be represented by a single unit cell with periodic boundary conditions. Identifying opposite edges makes the unit cell topologically equivalent to a torus, with $g=1$ and $\chi=0$, thus the net index of the current field must vanish. In a square lattice, positive circulating defects at the obstacles and cell centers are compensated by negative cross-defects at the inter-obstacle gaps [Fig.~\ref{rotlet}d]. The experimentally observed orbits around obstacles, voids, and gaps are therefore different manifestations of a single zero-charge defect network. The lattice geometry determines its local substructure, whereas the global index balance is fixed by the topology of the accessible space the boundary conditions. Changing the obstacle motif can reshape the defect network but cannot alter its local index except through defect creation, annihilation, or boundary crossing. Additional few-obstacle motifs, together with robustness to enhanced noise, lattice disorder, conservative drifts, and Brinkman-like hydrodynamic screening are discussed in the SM [Fig.~S2-S5].

\begin{figure*}[!ht]
\includegraphics[width=1.0\linewidth]{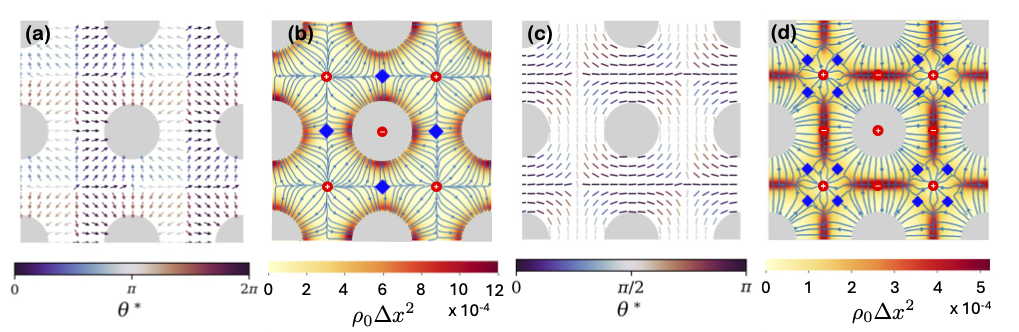}
\caption{
Obstacle-induced orientation fields and steady-state transport of translating singularities in a square obstacle lattice ($l_p=4.0 r_o$).
(a,c) Stable orientation fields selected by the obstacle-scattered flow for (a) a polar stokeslet and
(c) an apolar force dipole. Color denotes the local angle $\theta^\ast$, using $2\pi$- and $\pi$-periodic colormaps for the polar and nematic fields, respectively.
(b) For the stokeslet, the polar field projects onto a real-space drift; streamlines show the resulting stationary probability current over the scaled density $\rho_0 \Delta x^2$.
(d) For the force dipole, nematic symmetry cancels the polar advective current; streamlines instead show the active focusing flux obtained from the stationary state balance, again over $\rho_0 \Delta x^2$. Red symbols mark positive-index source- or sink-like defects, and blue diamonds mark negative-index saddle defects. Here $\Delta x$ is the grid spacing.
}
\label{together}
\end{figure*}

Rotlets provide the most direct connection between the topological organization of nonequilibrium currents and transport because, once the rotation axis and chirality are fixed, their state is specified solely by position $\mathbf{r}$. The obstacle-scattered flow therefore acts directly as a positional drift $\mathbf V(\mathbf r)$, generating a real-space probability current. Translating hydrodynamic singularities, such as stokeslets and force dipoles, instead require both position and orientation, $(\mathbf r,\theta)$; consequently, their dynamics naturally evolves in position--orientation space. The angular probability current is 
\begin{equation}
    J_\theta
    =
    \Omega^{\rm sc}(\mathbf r,\theta)P(\mathbf r,\theta,t)
    -
    D_r\partial_\theta P(\mathbf r,\theta,t) ,
    \label{eq:angular_current}
\end{equation}
where $P$ is the probability density, $D_r$ the rotational diffusion coefficient, and $\Omega^{\rm sc}=(1/2)[\nabla\times\mathbf u^{\rm sc}]_z$ the angular drift generated by the local vorticity of the obstacle-scattered flow $\mathbf u^{\rm sc}$ generated by a singularity at position $\mathbf r$ with orientation $\theta$, evaluated at the particle position. The lattice-selected orientations are the stable zeros of the angular drift, $\Omega^{\rm sc}(\mathbf r,\theta^\ast)=0$, with $\partial_\theta\Omega^{\rm sc}(\mathbf r,\theta^\ast)<0$, defining a stable orientation field $\theta^\ast(\mathbf r)$. When angular relaxation is faster than translating across one lattice spacing, $\tau_r < \tau_t \simeq l_p / V_0$ where $V_0$ is the particle speed, this field controls the transport.

We first consider the stokeslet, the simplest polar translating singularity. A point force $\mathbf F=F\hat{\mathbf p}$ generates the Oseen flow in an unbounded fluid~\cite{microhydrodynamics} [Fig.~S6], with transport direction $\hat{\mathbf p}=(\cos\theta,\sin\theta)$. In the lattice, obstacles scatter this primary flow [Fig.~S7]. We compute the resulting $\mathbf u^{\rm sc}$ using the regularized-stokeslet method~\cite{cortez2005} and obtain $\Omega^{\rm sc}$ from its local vorticity (see SM). The attracting zeros give the stable polar field $\hat{\mathbf p}^{\ast}(\mathbf r)=(\cos\theta^\ast(\mathbf r),\sin\theta^\ast(\mathbf r))$ shown in Fig.~\ref{together}a. In the slaving regime, this field sets the projected drift $\mathbf V(\mathbf r)=V_0\hat{\mathbf p}^{\ast}(\mathbf r)$, which determines the stationary density and current in Fig.~\ref{together}b.

Transport is organized by the singularities and winding indices of the stable polar field, rather than by the detailed magnitude of the scattered flow. In the square lattice, $q=+1$ aster defects occur at the obstacle centers, while $q=-1$ saddle defects lie at inter-obstacle gaps [Fig.~\ref{together}a], giving zero net index per unit cell. This is the same global balance as for rotlets, despite the different local defect morphology. Rotlet currents are chiral and tangential, $\mathbf W^\varphi\propto\hat{\mathbf z}\times\hat{\mathbf r}$, producing vortex-like $q=+1$ defects, whereas stokeslets scattering generates polar fields with aster-like defects. Because vortices and asters are related by a smooth local rotation of the local vector direction, $\hat{\boldsymbol{\varphi}}=R_{\pi/2}\hat{\mathbf r}$, their winding index is unchanged, and the same holds for the corresponding saddle defects. Thus, although a stokeslet translates even without obstacles, the lattice topologically organizes its allowed directions of motion, and the projected Fokker--Planck current converts this orientational topology into real-space active transport [Fig.~\ref{together}b].

Force dipoles provide the apolar counterpart to stokeslets. As the leading far-field representation of force-free swimmers~\cite{Drescher2011}, their flow is invariant under $\hat{\mathbf p}\to-\hat{\mathbf p}$ [Fig.~S6]. Under the quasi-2D confinement considered here, wall-induced alignment~\cite{spagnolie2012} restricts the relevant stable orientations to an in-plane nematic director field, $\hat{\mathbf n}^{\ast}\equiv-\hat{\mathbf n}^{\ast}$. Although this nematic symmetry permits half-integer defects, only integer-index singularities arise under the conditions considered here. We obtain $\hat{\mathbf n}^{\ast}$ from the obstacle-scattered flow generated by a dipole [Fig.~S7] by identifying the stable orientations of the local angular drift [Fig.~\ref{together}c]. Its apolar, $\pi$-periodic symmetry favors alignment parallel to the no-slip obstacle boundaries, producing a defect morphology distinct from the rotlet and stokeslet cases: positive vortex-like defects around the obstacles, a negative anti-aster at the central void, and the inter-obstacle gaps remain nearly uniformly aligned. These defects compensate to give zero net index under periodic boundary conditions.

Apolar symmetry changes how topology appears in transport. Unlike the stokeslet, whose polar orientation field projects directly onto a positional drift, a nematic director specifies an axis but not a polarity; choosing a global polar branch would introduce artificial discontinuities and break the intrinsic $\hat{\mathbf p}\to-\hat{\mathbf p}$ symmetry. We therefore simulate self-propelled particles whose orientations couple continuously to the nematic field, without imposing such polar branch (details in the SM). The resulting trajectories form closed, boundary-following orbits around individual obstacles [Fig.~S8], and probability accumulates at the inter-obstacle gaps where neighboring obstacle-bound orbits overlap [Fig.~\ref{together}d]. This density modulation arises not from a conservative potential but from active focusing by self-propulsion and nematic alignment. Because the two branches $\pm\hat{\mathbf n}^{\ast}$ are equivalent, the director field alone defines no net advective current along it; nevertheless, particles self-propel along one branch at a time, generating an active positional flux that is balanced in the stationary state by diffusion, $\mathbf J_{\rm act}=D_t\nabla\rho_0$. This focusing field has its own defect structure: inter-obstacle gaps form attracting nodes, while repelling nodes and saddles near central voids, obstacles, and lattice diagonals partition the unit cell into basins of obstacle-bound orbits. Transport therefore proceeds through noise-assisted hopping between topologically defined orbit basins rather than continuous advection along a polar current.

In summary, structured environments can organize active transport topologically through hydrodynamic scattering. For chiral rotlets, obstacle-scattered flows generate nonequilibrium probability currents whose integer-index defects organize spinner motion. For translating singularities, this organization first emerges in position--orientation space, where scattering selects stable polar or nematic orientation fields. Real-space transport then depends on the symmetry of these fields and the global index constraints imposed by the accessible space: polar stokeslets generates continuous advective currents, whereas apolar force dipoles form closed obstacle-following orbits whose overlap produces active focusing at inter-obstacle gaps. This focusing arises not from a conservative potential but from an active flux balanced by diffusion, revealing broken detailed balance even when nematic symmetry prevents a net polar advective current. Across these systems, the hydrodynamic singularity and obstacle geometry determine the local defect morphology, while the global index balance is fixed by the topology of the accessible space and its boundary conditions.

These results identify the topology of nonequilibrium currents as a robust organizing principle for active transport. Unlike topological transport in quantum or metamaterial systems, where robustness is typically associated with band invariants, the invariant here is encoded directly in the stochastic fields that transport or focus active particles. Transport pathways and focusing regions are therefore constrained by defect conservation rather than prescribed microscopic trajectories. Noise, disorder, and smooth geometric perturbations can displace these structures, but cannot remove them without defect creation, annihilation, or boundary crossing. Beyond winding numbers and index constraints, the connectivity of saddles, separatrices, and transport basins offers a natural next level of classification, distinguishing fields with identical total charge but different transport networks. This links active matter and topology at the level of stochastic transport itself, and provides a framework for designing active materials, microfluidic devices, and microrobotic systems whose macroscopic transport is controlled by current-field topology.

\begin{acknowledgments}
\noindent We thank J. del Pino for fruitful discussions and valuable comments on the manuscript. This work was supported by MCIN/AEI/10.13039/501100011033/ for all grants listed next: JVA for PID2022-139776NB-C64; JLA and LRA for PID2022-143010NB-I00; JVA, JLA and LRA for CEX2023-001316-M; all also supported by "ERDF A way of making Europe"; GG for PRE-2021-099492, also supported by "ESF Investing in your future". AAK is grateful to MISTI Spain for funding and to the Michael and Sonja Koerner chair for financial support as well.
\end{acknowledgments}

\nocite{*}

\bibliographystyle{apsrev4-2}
%

\end{document}